\documentclass[aps,prl,reprint]{revtex4-1}
\usepackage{amsmath}
\usepackage{graphicx}
\usepackage{color}
\usepackage{amssymb}
\usepackage{wasysym}
\usepackage{amsthm}
\usepackage{soul}
\usepackage{mathtools}   % loads �amsmath�
\usepackage[T1]{fontenc}
\input{epsf}

\begin{document}

\bibliographystyle{apsrev}

\renewcommand{\andname}{\ignorespaces}

\title{Hard competition: stabilizing the elusive biaxial nematic phase in suspensions of colloidal particles with extreme lengths}

\author{Simone Dussi$^{1,3,\dag}$, Nikos Tasios$^{1,\dag}$, Tara Drwenski$^{2,\dag}$, Ren\'e van Roij$^{2}$, Marjolein Dijkstra$^{1}$}
\affiliation{$^1$Soft Condensed Matter, Debye Institute for Nanomaterials Science, Utrecht University, Princetonplein 5, 3584 CC Utrecht, The Netherlands}
\affiliation{$^2$Institute for Theoretical Physics,  Utrecht University, Princetonplein 5, 3584 CC Utrecht, The Netherlands}
\email{m.dijkstra@uu.nl}
\altaffiliation{$^3$ Present address: Physical Chemistry and Soft Matter, Wageningen University, Stippeneng 4, 6708 WE, Wageningen, The Netherlands}
\thanks{$^{\dag}$~These authors contributed equally to this work.}

\begin{abstract}
We use computer simulations to study the existence and stability of a biaxial nematic $N_b$ phase in systems of hard polyhedral cuboids, triangular prisms, and rhombic platelets, characterized by a long ($L$), medium ($M$), and short ($S$) particle axis. For all three shape families, we find stable $N_b$ states provided the shape is not only close to the so-called dual shape with $M = \sqrt{LS}$ but also sufficiently anisotropic with $L/S>9,11,14, 23$ for rhombi, prisms, and cuboids, respectively, corresponding to anisotropies not considered before.  Surprisingly, a direct isotropic-$N_b$ transition does not occur in these systems due to a destabilization of $N_b$ by a smectic (for cuboids and prisms) or a columnar (for platelets) phase at small $L/S$, or by an intervening uniaxial nematic phase at large $L/S$. 
Our results are confirmed by a density functional theory provided the third virial coefficient is included and a continuous rather than a discrete (Zwanzig) set of particle orientations is taken into account.

\medskip

%82.70.Dd Colloids
%64.70.Md Transitions in liquid crystals
%61.30.Cz Molecular and microscopic models and theories of liquid crystal structure
%61.30.-v Liquid crystals

\end{abstract}

\maketitle
\noindent
Anisotropic molecules, viruses, wormlike micelles, and suspended nanoparticles can form liquid-crystal phases which exhibit long-range order of the particle orientations, possibly combined with some degree of positional order~\cite{onsager1949,maier1959einfache,degennes1993}. The simplest liquid-crystal state is the homogeneous {\em nematic} phase, which exhibits only orientational order. However, this simplicity is only apparent. For instance, the microscopic origin of the chiral nematic (cholesteric) phase and the twist-bend nematic phase is still not well understood, even though recent advances in particle synthesis~\cite{xia2009,gong2012,zhang2012}, microscopy techniques, and computer simulation~\cite{escobedo2011,damasceno2012,mederos2014,kolli2014softmatter,greco2015entropy,kuhnhold2016isotropic,ruuvzivcka2016simulating, dijkstra2015,dussi2016} have provided new insights. 

\begin{figure}[h!t]
\begin{center}
\includegraphics[width=0.45\textwidth]{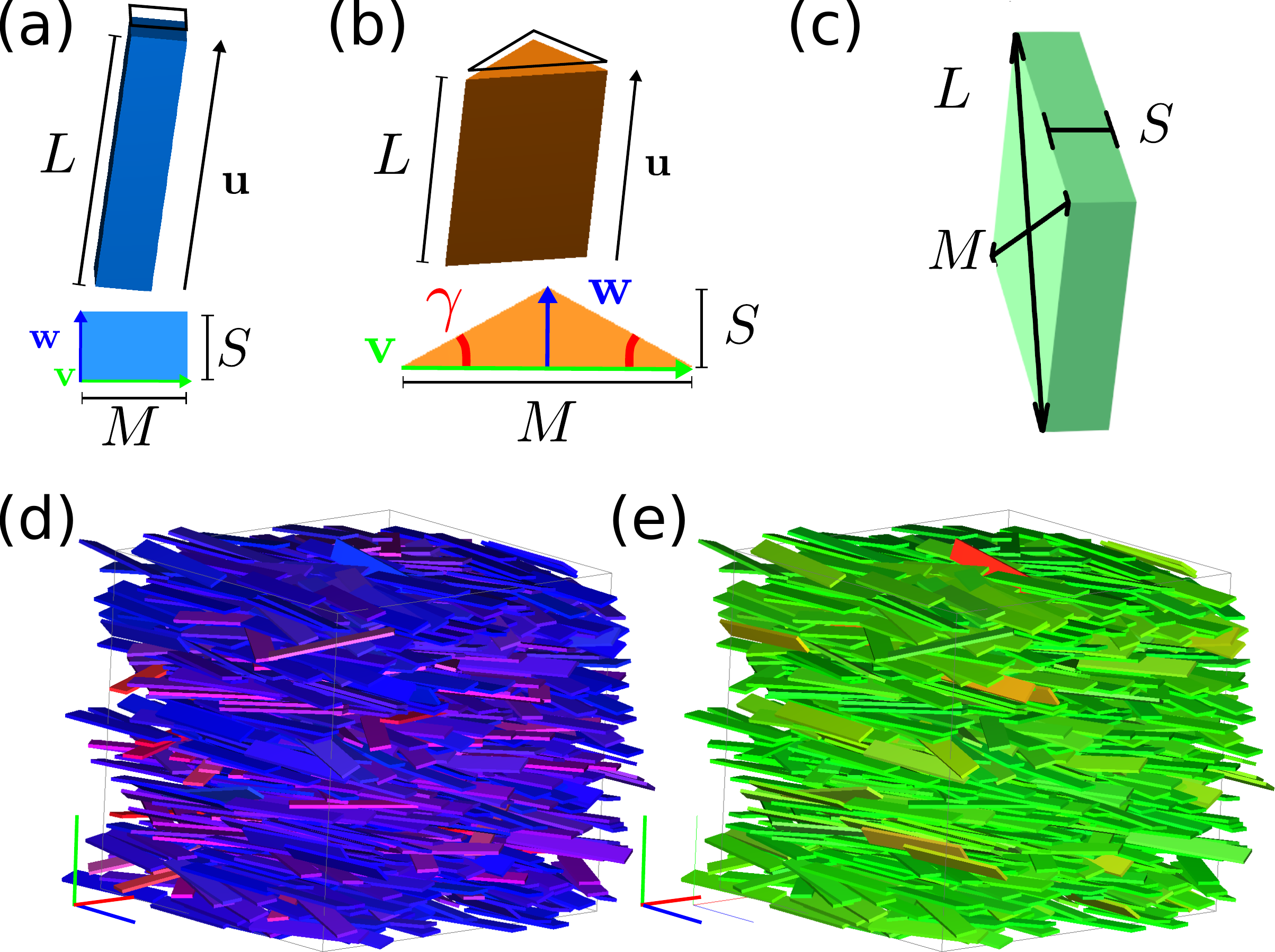}
\end{center}
\caption{The shape families of  (a) cuboids, (b) triangular prisms, and (c) rhombic platelets with particle axes $L$, $M$, and $S$. Representative simulation snapshots (d) and (e) of a biaxial nematic  $N_b$ phase composed of hard cuboids with $L/S=35$ and $M/S\simeq5.9$ ($\nu=0$) at packing fraction $\eta\simeq0.3$ of the same configuration with the particles colored according to the orientation of their (d) long or (e) short particle axis. Snapshots of an $N_b$ phase for triangular prisms and rhombic platelets are shown in the Supplemental Material~\cite{SM}. }
\label{fig1}
\end{figure}

A problem with an even longer history concerns the existence of a stable biaxial nematic $N_b$ phase, which exhibits two optical axes in contrast to the ordinary uniaxial nematic phase that displays only a single optical axis. Biaxial nematic phases have long held promise for applications in novel opto-electronic devices, but their limited window of thermodynamic stability (and for a long time even their very existence) has been of great concern. The theoretical prediction of the existence of the $N_b$ phase goes back to the 1970's~\cite{freiser1970,straley1974}, and first claims of its experimental observation in a micellar system date back to 1980~\cite{yu1980}. In 2004, novel experiments on different molecular thermotropic systems again claimed to observe the $N_b$~\cite{luckhurst2004,severing2004,madsen2004,acharya2004}, which in the meantime was also observed in computer simulations of attractive particles~\cite{berardi2000,berardi2008}. More recently, an $N_b$ phase was observed in colloidal dispersions of purely repulsive board-like particles in 2009~\cite{vandenpol2009}, where the stability was argued to stem from polydispersity that prevents the system from forming a smectic phase~\cite{belli2011}. This finding in an entropy-dominated system appears to be consistent with the observation of a stable $N_b$ phase in early simulations of hard biaxial ellipsoids~\cite{allen1990,camp1997}, which do not exhibit a smectic phase either~\cite{frenkel1984phase}. Interestingly, however, recent simulations of hard spheroplatelets (with a stable smectic phase in their phase diagram) also revealed a stable $N_b$ phase~\cite{peroukidis2013}, whereas ostensibly similarly shaped cuboidal particles do not~\cite{cuetos2017}. On top of this confusing situation comes an unsettled issue regarding the topology of the phase diagram, in particular whether a prolate ($N_+$) or oblate ($N_-$) uniaxial nematic phase intervenes the isotropic ($I$) and $N_b$ phase or whether a direct $I-N_b$ phase transition is possible.  According to early theoretical studies the density-shape representation of the phase diagram exhibits a cusp-like feature where a rod-like regime with $I-N_+$ coexistence and a plate-like regime with $I-N_-$ coexistence merge at the so-called dual shape into a single multi-critical point with a direct $I-N_b$ phase transition~\cite{mulder1989,taylor1991}. More recent Landau-type theories, however, also allow for other scenarios either with or without a direct $I-N_b$ transition~\cite{longa2005luckhurst,allender2008landau,mukherjee2009new,luckhurst2015}.

In this Letter, we will settle the issue of the existence and stability of the biaxial nematic phase in entropy-driven systems by performing computer simulations of three different families of hard biaxial particles, extending the range of shape parameters to anisotropies much beyond hitherto considered. We will see that the $N_b$ phase can be stable close to the dual shape, as expected, but only if the particle anisotropy exceeds a critical value which, surprisingly, varies significantly between the different particle families. In fact, we find that strong  competition with the $N_b$ phase does not only come from the smectic phase at high densities, but also from the uniaxial $N_+$ and  $N_-$ phases at relatively low densities, such that a direct $I-N_b$ phase transition does not exist due to an intervening uniaxial nematic phase. The absence of this direct $I-N_b$ transition in our simulations is confirmed by a third-virial density functional theory with continuous rather than discrete orientations of the particles. For less anisotropic shapes, both uniaxial and biaxial nematic phases are absent from our simulations, and a direct $I$ to a positionally-ordered liquid crystal phase (smectic $Sm_+$ for cuboids and prisms, columnar $Col$ for platelets) transition is observed. 

We consider the three different families of hard particles shown in Fig.~\ref{fig1}: (a) cuboids, (b) triangular prisms, and (c) rhombic platelets, all characterized by long ($L$), medium ($M$), and short ($S$) particle axes that give rise to the dimensionless particle length $L^*\equiv L/S$, particle width $M^*=M/S$,  and the particle shape parameter $\nu=S/M-M/L\in[-1,1]$ --only two of which are needed to fully characterize the shape for a given family. Rod-like shapes with $\nu>0$ are expected to feature a prolate $N_+$ phase in their phase diagram, and plate-like shapes with $\nu<0$ are expected to form an oblate $N_-$ phase. The case $\nu=0$ (or $M=\sqrt{LS}$ or $L^*=(M^*)^2$) refers to the dual shape where biaxial nematic phases could be expected~\cite{mulder1989}. We determine the phase behavior of more than 100, 60, and 20 members, as characterized by different $L^*$ and $M^*$, of the cuboid, rhombic platelet, and triangular prism family, respectively, all as a function of packing fraction $\eta$, by performing Monte Carlo and Event-Driven Molecular Dynamics simulations of systems consisting of thousands of identical particles.   We show that all three families have members that exhibit a stable $N_b$ phase such as illustrated for cuboids with $L^*=35$ and $M^*\simeq5.9$ ($\nu=0$) at $\eta\simeq0.3$ in Fig.~\ref{fig1}, where the same configuration is shown twice with a color coding representing the alignment of the long (d) and short (e) particle axes. We distinguish the different liquid-crystalline phases with a variety of scalar and tensorial order parameters~\cite{SM}.

\begin{figure}[h!t]
\begin{center}
\includegraphics[width=0.45\textwidth]{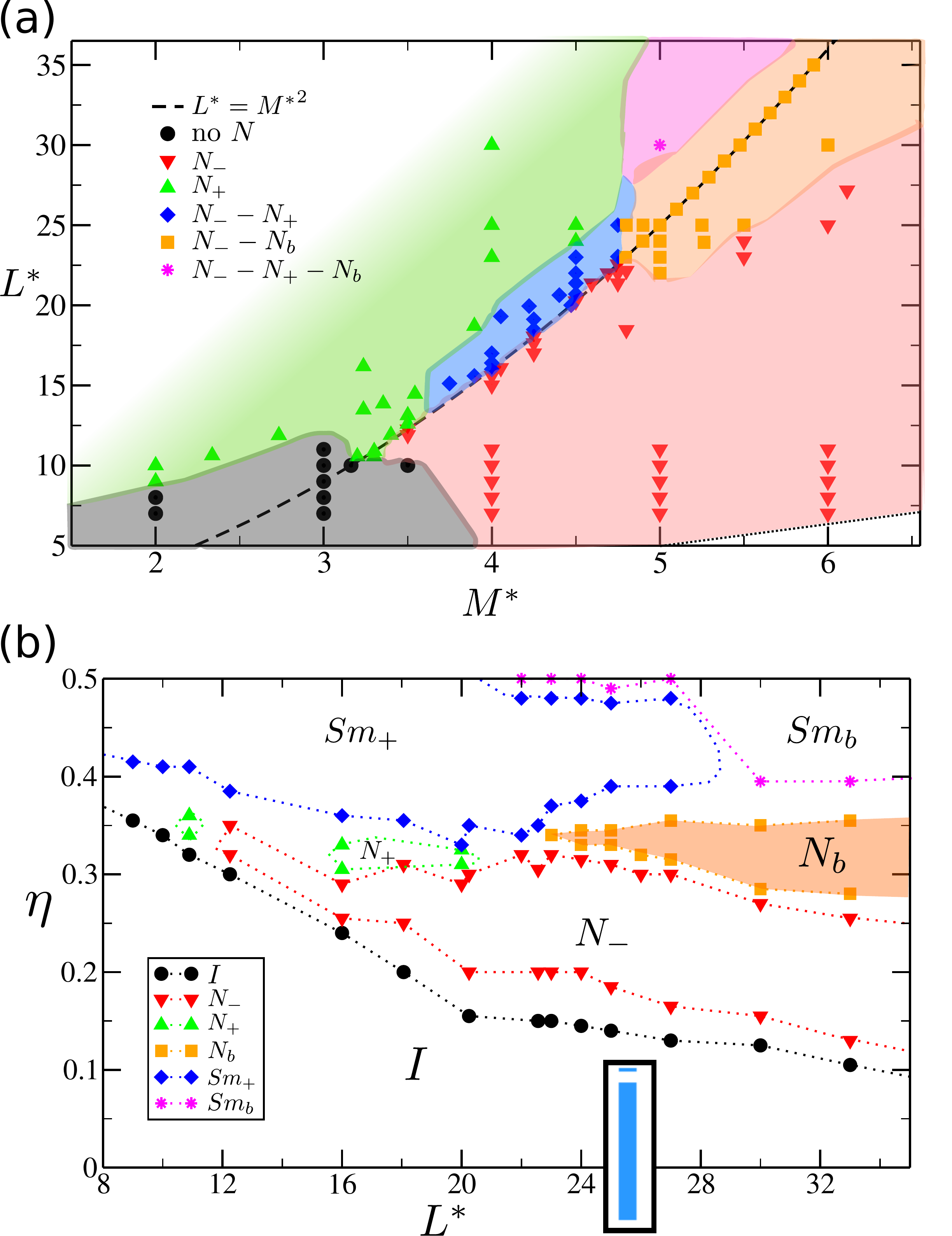}
\end{center}
\caption{(a) Meta-phase diagram of hard cuboids in the dimensionless particle width and length plane spanned by $(M^*,L^*)$. Symbols denote simulated particle shapes, the dashed line indicates dual shapes $L^*=M^{*2}$, and different colors denote different sequences of oblate, prolate, and biaxial nematic phases $N_-$,  $N_+$, and $N_b$, respectively, upon increasing the density.  (b) Phase diagram of dual-shaped cuboids in the packing fraction $L^*-\eta$ representation, featuring an additional isotropic phase $I$ as well as prolate and biaxial smectic phases $Sm_+$ and $Sm_b$.   The inset shows cross-sections of a dual-shaped cuboid with $L^*=24$.}
\label{fig2}
\end{figure}

\begin{figure*}[h!t]
\begin{center}
\includegraphics[width=0.95\textwidth]{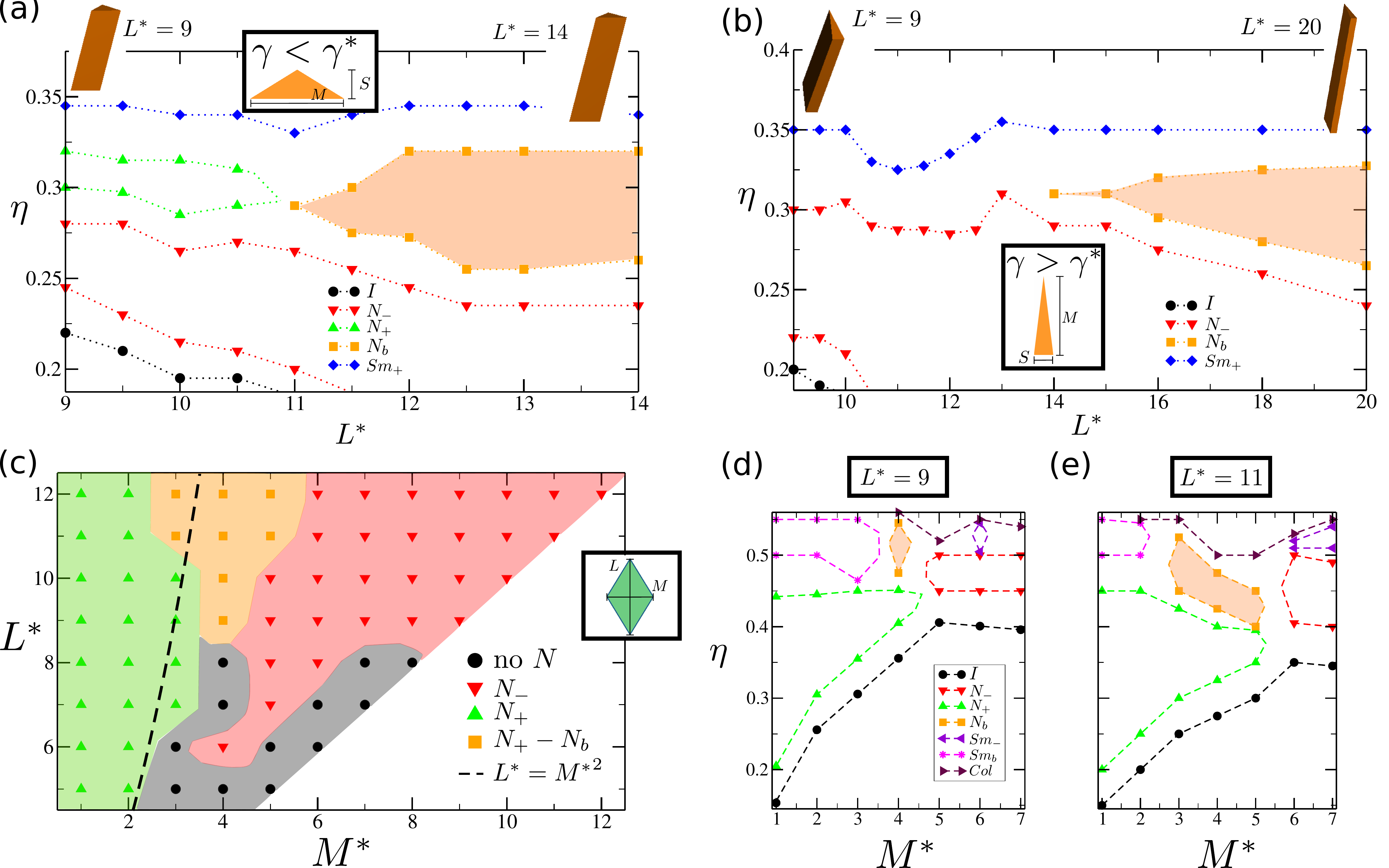}
\end{center}
\caption{Phase diagram of dual-shaped triangular prisms in the $L^*-\eta$ representation for (a) $\gamma<\gamma^* = \pi/3$ and (b) $\gamma>\gamma^*$, see inset and text. The dashed lines denote the binodals.  (c)  Meta-phase diagram of hard rhombic platelets in the  $(M^*,L^*)$ plane. Symbols denote simulated particle shapes, the dashed line indicates dual shapes $L^*=M^{*2}$, and different colors denote different sequences of  nematic phases  upon increasing the density. Phase diagram of hard rhombic platelets in the $M^*-\eta$ representation for (d) $L^*=9$ and (e) $L^*=11$, featuring an oblate smectic $Sm_-$ phase and columnar $Col$ phase.}
\label{fig3}
\end{figure*}

In Fig.~\ref{fig2}(a) we show the meta-phase diagram of the family of cuboids in the $(M^*,L^*)$ plane, where the symbols represent actually simulated particle shapes and the dashed line corresponds to the dual shape ($L^*=M^{*2}$ or $\nu=0$). We identify different regions according to the sequence of nematic phases as observed from low to high density beyond the isotropic phase; for clarity we disregard the high-density smectic, columnar, and crystal phases. The black (circles) region corresponds to particle shapes that are not sufficiently anisotropic to form any nematic phase due to a direct transition from an isotropic ($I$) to a prolate smectic ($Sm_+$) phase, in accordance with recent findings~\cite{cuetos2017}. As expected, rod-like cuboids with $L^*\gg M^{*2}$, i.e., far above the dashed line, form $N_+$ phases (green, upsided triangles),  and plate-like cuboids with $L^*\ll M^{*2}$, i.e., far below the dashed line, form $N_-$ phases (red, downsided triangles). The phase sequences close to the dual shape are more interesting. At intermediate $L^* \sim 16-23$ and $M^* \sim 4-5$,  in the rod-like regime above the dashed line, we find a plate-like nematic $N_-$ phase that remarkably intervenes between the $I$ and the $N_+$ phase  (blue, diamonds).   Surprisingly, a biaxial nematic phase only appears in Fig.~\ref{fig2}(a) for cuboids close to the dual shape when particle anisotropies are as extreme as $L^*>23$ and $M^*>5$, again with an intervening $N_-$ phase (orange, squares). Quite deep in the rod-like regime, $L^*>30$ and $M^*>5$, we even observe an intricate $N_- - N_+ - N_b $ phase sequence (pink, asterisks). Hence, the nematic phase in coexistence with the isotropic phase is $N_-$ in a substantial region even where $L^*> M^{*2}$ (or $\nu>0$). No direct $I-N_b$ transition is found for cuboidal particles. This conclusion is even more apparent in Fig.~\ref{fig2}(b), where we present the phase diagram of dual-shaped cuboids in the $L^*-\eta$ representation, where the lines denote the approximate binodals. For $L^*<11$, we observe the strongly first-order phase transition from the $I$ to a prolate smectic ($Sm_+$) phase.  Furthermore Fig.~\ref{fig2}(b) reveals for $L^*>12$ that the coexistence with the isotropic phase is solely with the  $N_-$ phase, which is stable for a surprisingly large density regime before it  transforms into an $N_b$ phase for $L^*>23$ and into $N_+$ for  $16<L^*<20$.  The $N_b$ phase is seen to be stable for a wider range of $\eta$ upon increasing $L^*$.  Compressing the $N_b$ to the smectic phase leads to biaxial order being lost unless $L^*>28$. 

Our present study extends significantly beyond previous simulations on cuboids with $M^*=1$ and $0.125 \leq L^* \leq 5$, where no biaxial order was found nor expected~\cite{john2008}. In the search for a biaxial nematic phase, Cuetos \emph{et al.}~\cite{cuetos2017} recently performed extensive simulations of more anisotropic cuboids with $L^*=9, 12$ and $1 \leq M^* \leq  L^*$, but an $N_b$ phase was not observed, consistent with our finding that  $L^*>23$ is required. Interestingly, dual-shaped cuboids with $L^*=23$ have $L/M=4.8$, which is large enough for uniaxial rods to form a stable nematic phase~\cite{bolhuis1997tracing}, whereas $L^*=12$ only yields $L/M=3.5$ which does not suffice.

We now turn our attention to the phase behavior of triangular prisms with a dual shape ($L^*=M^{*2}$) and an isosceles triangular base with a base angle  $\gamma$ as defined in Fig.~\ref{fig1}(b). 
In Fig.~\ref{fig3} we present the phase diagram of these particles in the $L^*-\eta$ representation, in (a) for $\gamma<\gamma^*$  and in (b) for $\gamma>\gamma^*$, where $\gamma^* = \pi/3$ is the cross-over angle between one large and two small sides of the triangular base for $\gamma<\gamma^*$ to vice versa for $\gamma>\gamma^*$ (see insets).
Both phase diagrams display  an  $N_--N_b$ phase sequence  for sufficiently large particle anisotropies $L^* \geq 11$ in (a) and $L^* \geq 14$ (b), i.e., without any direct $I-N_b$ transition due to an intervening $N_-$ phase. For smaller $L^*$, we observe an $N_--N_+-Sm_+$ phase sequence in (a) whereas this intervening $N_+$ phase is absent in (b).

Next we consider in Fig.~\ref{fig3}(c) the meta-phase diagram of the family of rhombic platelets in the $M^*-L^*$ plane (with $L^*>M^*$), with a focus on the sequence of nematic states upon increasing the density. The dual shape is represented by the dashed line. We again observe $N_+$ for $L^*\gg M^{*2}$ and $N_-$ for $L^*\ll M^{*2}$, but now with prolate order invading the region $\nu<0$. A striking feature is the appearance of an $I-N_+-N_b$ sequence (orange, squares), not only above the dashed line but also below it in the ``plate-like'' regime.  As for cuboids and triangular prisms, we also find that the family of rhombic platelets does not display a direct $I-N_b$ transition. This is also evident from the $M^*-\eta$ phase diagrams for $L^*=9$ and $L^*=11$ shown in Fig.~\ref{fig3}(d) and (e), respectively. Fig.~\ref{fig3}(d) and (e) also show that the $N_b$ phase transforms into a columnar phase $Col$ upon increasing the density instead of a smectic phase in the case of cuboids and triangular prisms. 

\begin{figure}[h!t]
\begin{center}
\includegraphics[width=0.45\textwidth]{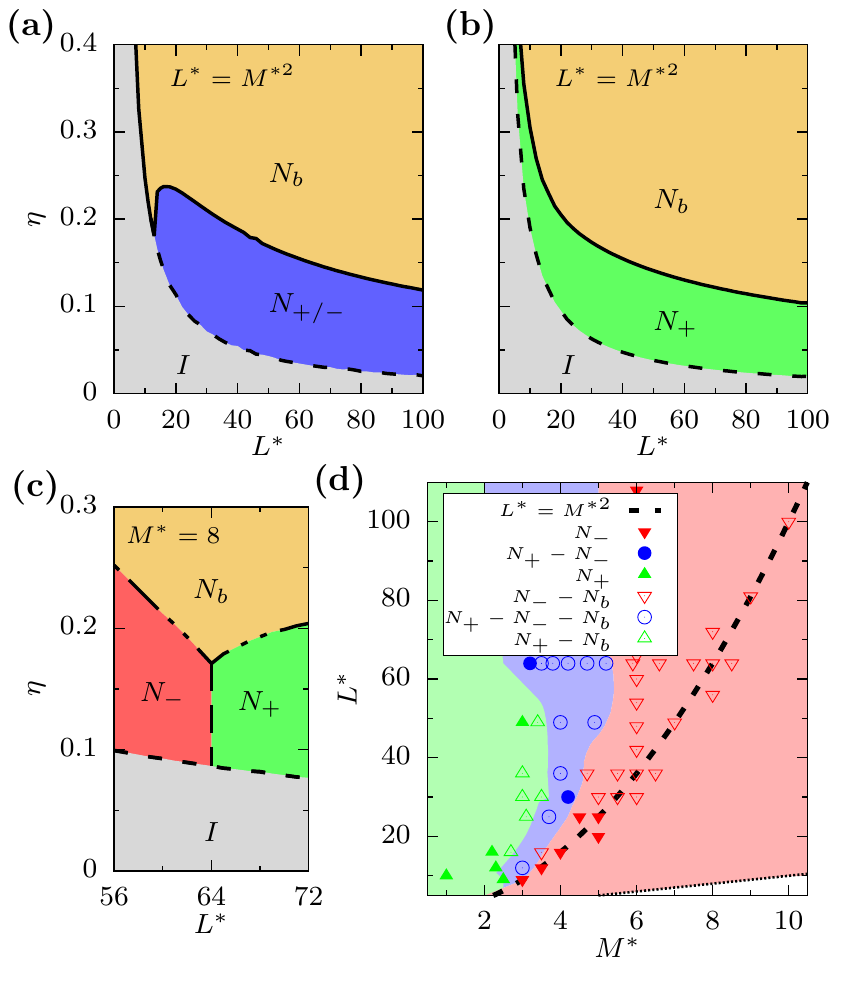}
\end{center}
\caption{Phase diagram for dual-shaped cuboids in the $L^*-\eta$ plane for a Zwanzig model within the (a) second- and (b) third-virial theory. (c) Phase diagram for $M^*=8$ cuboids in the $L^*-\eta$ plane within the full-orientation second-virial theory. (d) Meta-phase diagram of cuboids in the ($M^*$,$L^*$) plane for continuum orientations within a full third-virial theory. Different symbols denote different nematic phase sequences from low to high packing fraction (up to $\eta = 0.6$ for $L^*<50$ and up to $\eta=0.4$ for $L^*\geq50$), neglecting spatially ordered phases.}
\label{fig4}
\end{figure}

In order to shed further light on the subtleties involved in the absence of a direct $I-N_b$ transition due to the intervening $N_-$ phase in the case of biaxial cuboids, we also employed density-functional theory for these particle shapes. We consider both a second- and a third-virial theory, not only for the Zwanzig model with six discrete particle orientations~\cite{belli2011,martinez2011} but also for  cuboids with a continuum of particle orientations treated within an expansion into Wigner matrices~\cite{SM}. Within a second-virial theory there is perfect symmetry of prolate and oblate ordering 
with respect to the dual shape. Contrary to Mulder's conclusion on the basis of a bifurcation analysis~\cite{mulder1989} our numerical free-energy minimizations yield {\em no} direct $I-N_b$ transition, at least for dual-shaped Zwanzig cuboids with $L^*>11$, but rather an $I-N_+/N_- -N_b$ sequence, where the $N_+/N_-$ phases have identical free energies within the second-virial theory, see Fig.~\ref{fig4}(a). However, in agreement with~\cite{belli2011} we do find a direct $I-N_b$ transition for $L^*\leq 11$. In the third-virial theory for the Zwanzig model, the $N_+/N_-$ symmetry is broken in favor of $N_+$ at $\nu=0$ as shown in Fig.~\ref{fig4}(b), in agreement with the Fundamental Measure Theory results of Ref.~\cite{martinez2011}, but not in agreement with our simulations of the continuum model, where $N_-$ is favored.  
Applying the second-virial theory to the continuum model restores the symmetry about $\nu=0$ as  illustrated in the phase diagram of cuboids with $M^*=8$  in the $L^*-\eta$ representation in Fig.~\ref{fig4}(c), yet again in disagreement with our simulations, where the symmetry is broken. Only for a continuum of orientations combined with the third-virial approximation is qualitative agreement with the simulations obtained, since then the phase sequence $I-N_--N_b$ appears for $\nu=0$. The resulting meta-phase diagram of the family of hard cuboids in the $(M^*,L^*)$ plane is shown in Fig.~\ref{fig4}(d), which clearly shows that the $N_-$ phase invades the rod-like regime above the dashed line that represents the dual shape. 

In conclusion, for three large families of biaxial particle shapes we have shown that the $N_b$ phase can be stable close to the dual shape, but only if the particle shape is sufficiently anisotropic with $L^*>L^*_{min}=9, 11, 14, 23$ for rhombi, prisms with angle $\gamma<\pi/3$ and $\gamma>\pi/3$, and cuboids, respectively. Interestingly, for dual-shaped spheroplatelets that closely resemble our cuboids at first sight, recent simulations~\cite{peroukidis2013,peroukidis2013b} revealed an $N_b$ phase for $L^*>9$ rather than $L^*>23$ for cuboids.  This trend  in which particle shapes with triangular, rounded, or rhombic cross-sections have smaller $L^*_{min}$ than cuboids suggests that the 2D packing efficiency of the particle cross-section in a smectic layer largely determines $L^*_{min}$~\cite{taylor1991,cuetos2017,martinez2008,marechal2017}. Rhombic platelets with their minimal $L^*_{min}$ are indeed ideal candidates for observing a $N_b$ phase. Finally, none of the members of the different shape families exhibit a direct $I-N_b$ transition  due to the surprising stability of an intervening uniaxial nematic phase. Future experimental efforts towards the self-assembly of biaxial colloids will benefit from these results, with the knowledge that imperfections, such as roundness, facet corners, or polydispersity in particle shape, can facilitate the self-assembly of biaxial nematic phases.

\smallskip   

This work is part of the D-ITP consortium, a program of the Netherlands Organization for Scientific Research (NWO) that is funded by the Dutch Ministry of Education, Culture and Science (OCW). We acknowledge financial support from an NWO-ECHO and an NWO-VICI grant. We thank Bela Mulder for helpful discussions and Oleg Lavrentovich for useful comments about experiments on thermotropic systems.

% \input{biaxial_PRL_main.bbl}

% \bibliography{biaxial_refs}

%%%%%%%%%% Merge with supplemental materials %%%%%%%%%%
% \pagebreak
\clearpage

\onecolumngrid
\begin{center}
\textbf{\large Supplemental Material \\}
\end{center}
\twocolumngrid

%%%%%%%%%% Merge with supplemental materials %%%%%%%%%%
%%%%%%%%%% Prefix a "S" to all equations, figures, tables and reset the counter %%%%%%%%%%
\setcounter{equation}{0}
\setcounter{figure}{0}
\setcounter{table}{0}
\setcounter{page}{1}
\makeatletter
\renewcommand{\theequation}{S\arabic{equation}}
\renewcommand{\thefigure}{S\arabic{figure}}
% \renewcommand{\bibnumfmt}[1]{[S#1]}
% \renewcommand{\citenumfont}[1]{S#1}
%%%%%%%%%% Prefix a "S" to all equations, figures, tables and reset the counter %%%%%%%%%%

\newcommand\ddfrac[2]{\frac{\displaystyle #1}{\displaystyle #2}}
\newcommand{\D}{\mathcal{D}}
\renewcommand{\andname}{\ignorespaces}

% \title{Supplemental Material \\ Entropic Stabilization of Biaxial Nematics: a Competition with Uniaxial and Positional Order for Extremely Long Particles}
% \author{Simone Dussi$^{1,3,\dag}$, Nikos Tasios$^{1,\dag}$, Tara Drwenski$^{2,\dag}$, Ren\'e van Roij$^{2}$, Marjolein Dijkstra$^{1}$}
% \affiliation{$^1$Soft Condensed Matter, Debye Institute for Nanomaterials Science, Utrecht University, Princetonplein 5, 3584 CC Utrecht, The Netherlands}
% \affiliation{$^2$Institute for Theoretical Physics,  Utrecht University, Princetonplein 5, 3584 CC Utrecht, The Netherlands}
% \email{m.dijkstra@uu.nl}
% \altaffiliation{$^3$ Present address: Physical Chemistry and Soft Matter, Wageningen University, Stippeneng 4, 6708 WE, Wageningen, The Netherlands}
% \thanks{$^{\dag}$~These authors contributed equally to this work.}
% \maketitle
\noindent

\section{Simulation methods}
To study the phase behavior of polyhedral hard rods (either cuboids or triangular prisms) we employed standard Monte Carlo (MC) simulations either in the $NPT$ or $NVT$ ensemble. System sizes range from $N \simeq2000$ to $N\simeq5000 $ and several million MC steps are performed before obtaining equilibrated configurations. For $NVT$-MC simulations, each MC step consists on average of $N/2$ attempts of translating a random particle and $N/2$ attempts of rotating a random particle. In the case of $NPT$-MC simulations, an additional attempt is performed at each step in order to either scale isotropically the box volume or change only one edge of the cuboidal simulation box. Equilibrium average density, order parameters and diffraction patterns are calculated based on around one hundred equilibrated configurations. For systems of rhombic particles, cluster moves are used in $NPT$-MC to perform volume changes move~\cite{tasios2017,tasios_thesis}. Particles interact only via a hard-core potential and overlaps are detected using algorithms based either on triangular-triangular intersection-detection, using the RAPID library~\cite{rapid}, or based on the GJK algorithm~\cite{gjk,gino}, depending on the particle model. In addition to MC simulations, rhombic platelets (and some selected cases of cuboids) are simulated with state-of-the-art Event-Driven Molecular Dynamics (EDMD)~\cite{tasios_thesis}. The GJK overlap-detection algorithm is combined with conservative advancement and near-neighbor list to efficiently simulate around $N \sim 2\cdot10^3$ rhombic particles in the $NVT$ ensemble. The moment of inertia and the mass of all particles is set to 1 and the system is simulated for over $10^4 \tau$, where $\tau=v_p^{1/3}/v_0$ is the reduced time unit, with $v_p$ particle volume and $v_0$ the initial velocity of each particle (velocities are initialized by using random unit vectors whereas angular velocities are initially set to zero). The equilibrium pressure is calculated from the particle collisions in the equilibrated configurations.

\section{Order parameters and phase identification}
To quantify the orientational and positional order in the system we use several order parameters.
The nematic order parameters for the biaxial particle models considered here are obtained by first constructing the following tensors
\begin{equation}
\mathcal{Q^{\mathbf{\hat{a}}}_{\alpha\beta}} = \frac{1}{N} \sum_{i=1}^N \left[ \frac{3}{2} \mathbf{\hat{a}}_{i\alpha} \mathbf{\hat{a}}_{i\beta}-\frac{{\delta}_{\alpha\beta}} {2} \right],
\end{equation}
where $\alpha, \beta=x,y,z$ component and $\mathbf{\hat{a}}=\mathbf{\hat{u}},\mathbf{\hat{v}},\mathbf{\hat{w}}$ denotes the three symmetry axes of the particle (see also Fig. 1 of the main text) and where $N$ is the number of particles and $\delta_{\alpha \beta}$ is the Kronecker delta. By diagonalizing each of these tensors we obtain three eigenvalues $\lambda^{+}_{a} \geq \lambda^{0}_{a} \geq \lambda^{-}_{a}$. We identify the (scalar) order parameter associated with the nematic order of the axis $\mathbf{\hat{a}}$ as the maximum of these eigenvalues: $S^{\mathbf{\hat{a}}} \equiv \lambda^{+}_{a}$. The corresponding eigenvector is the nematic director $\mathbf{\hat{n}}_\mathbf{\hat{a}}$.
These order parameters are used to distinguish between oblate and prolate nematic phases, and only partially for biaxial nematic phases. In fact, to precisely quantify the degree of (macroscopic) biaxial alignment of a nematic phase an additional (scalar) order parameter $\mathcal{B}$ is employed. Notice that different notations and slightly different approaches are employed to calculate the biaxial order parameter in computer simulations~\cite{allen1990,camp1997,rosso2007,berardi2008,peroukidis2013b}. We follow the procedure in Refs.~\cite{allen1990,camp1997} that consists in first identifying an appropriate orthonormal basis for the laboratory reference frame that is aligned with the two main directions of the biaxial phase. For each configuration, we identify the largest $S^{ \mathbf{\hat{a}}}$ and we define the $z$-axis of the laboratory reference frame as  $\mathbf{\hat{Z}} \equiv   \mathbf{\hat{n}}_\mathbf{\hat{a}}$, with $\mathbf{\hat{a}}$ the principle main axis of the particle. Then, we identify the second largest nematic order parameter $S^{ \mathbf{\hat{b}}}$ and we define the second axis of the laboratory reference frame as $\mathbf{\hat{Y}} \equiv \mathbf{\hat{n}}_{\mathbf{\hat{b}}} - (\mathbf{\hat{n}}_{\mathbf{\hat{b}}} \cdot \mathbf{\hat{Z}}) \mathbf{\hat{Z}} \simeq \mathbf{\hat{n}}_{\mathbf{\hat{b}}}$. Analogously, we define the third axis of the laboratory frame by orthogonalizing the third nematic director: $\mathbf{\hat{X}} \equiv \mathbf{\hat{n}}_{\mathbf{\hat{c}}} - (\mathbf{\hat{n}}_{\mathbf{\hat{c}}} \cdot \mathbf{\hat{Z}}) \mathbf{\hat{Z}} - (\mathbf{\hat{n}}_{\mathbf{\hat{c}}} \cdot \mathbf{\hat{Y}}) \mathbf{\hat{Y}}$, with $\mathbf{\hat{c}}$ the third symmetry axis of the particle. Finally, we compute 
\begin{equation}
\mathcal{B}=\frac{1}{3} \left( \mathbf{\hat{Y}} \cdot \mathcal{Q}^{ \mathbf{\hat{b}}} \cdot \mathbf{\hat{Y}} + \mathbf{\hat{X}} \cdot \mathcal{Q}^{ \mathbf{\hat{c}}} \cdot \mathbf{\hat{X}} - \mathbf{\hat{Y}} \cdot \mathcal{Q}^{ \mathbf{\hat{c}}} \cdot \mathbf{\hat{Y}} - \mathbf{\hat{X}} \cdot \mathcal{Q}^{ \mathbf{\hat{b}}} \cdot \mathbf{\hat{X}} \right)  ,
\end{equation}
where $\mathcal{B}$ is normalized such that it ranges from 0 to 1. Low values of $\mathcal{B}$ correspond to an isotropic phase or to a uniaxial phase and high values to a biaxial phase. In Refs.~\cite{peroukidis2013,peroukidis2013b} a biaxial nematic phase is further classified in $N_{b-}$ and $N_{b+}$, depending on the leading uniaxial order parameter. The authors observed that $N_{b-}$ is always formed at lower densities than $N_{b+}$, which indicates the preference for oblate order, in agreement with our results. For simplicity, we avoided this additional classification.

To identify the phase transition to a positionally ordered phase we generalize an order parameter that, for example, is often used to identify smectic phases of spherocylinders:
\begin{equation}
\tau^{\mathbf{\hat{a}}} = \max_{l} \left| \sum_{j=1}^N \exp \left( \frac{2\pi}{l} i \mathbf{r}_j \cdot \mathbf{\hat{n}}_\mathbf{\hat{a}} \right) \right| ,
\end{equation}
where $l$ is a real number, $\mathbf{r}_j$ denotes the position of particle $j$ and as before $\mathbf{\hat{n}}_\mathbf{\hat{a}}$ indicates the nematic director associated to the axis $\mathbf{\hat{a}}$. A large $\tau^{\mathbf{\hat{a}}}$ indicates one-dimensional positional order (layering) associated to the particle axis $\mathbf{\hat{a}}$. If only one of these order parameters is significantly larger than zero (typically $>0.4$), a smectic phase ($Sm_+$, $Sm_-$ or $Sm_b$ depending on which particle axis is aligned and if the biaxial order parameter is large) is identified. Two $\tau^{\mathbf{\hat{a}}}>0$ correspond to a columnar phase and three $\tau^{\mathbf{\hat{a}}}>0$ to a crystal phase. In addition, the positionally ordered phases are also identified by checking the (projected) diffraction patterns. In particular, the particle positions are projected on the plane defined by the two smallest nematic directors and subsequently we calculate the Fourier transform of a two-dimensional histogram of the projected positions. 

\begin{figure*}[h!t]
\begin{center}
\includegraphics[width=0.8\textwidth]{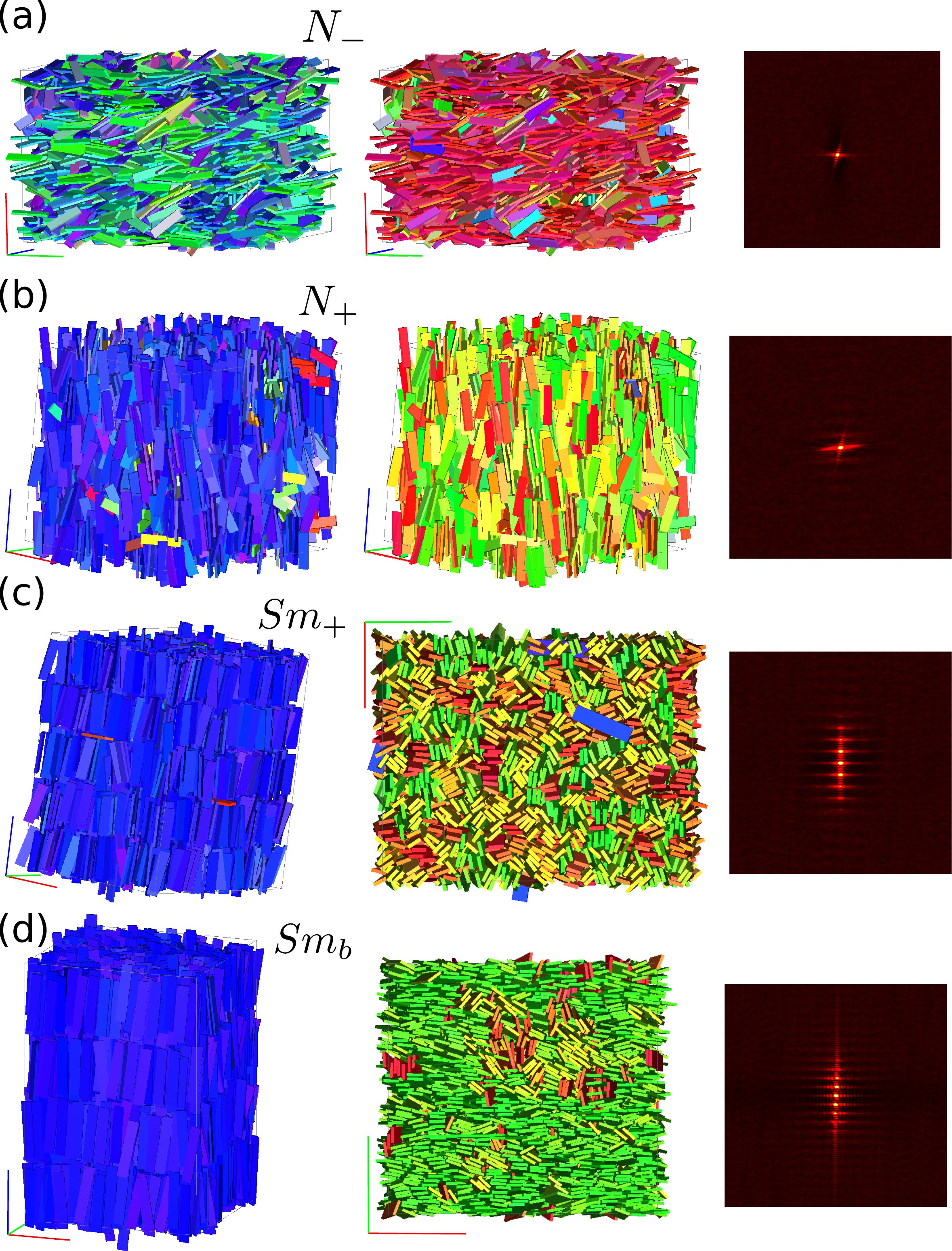}
\end{center}
\caption{Representative snapshots and corresponding diffraction patterns for cuboids forming (a) $N_-$ ($L^*=16$, $M^*=4$, $\beta P v_p=2.50$), (b) $N_+$ ($L^*=16$, $M^*=4$, $\beta P v_p=3.0$), (c) $Sm_+$ ($L^*=16$, $M^*=4$, $\beta P v_p=4.0$), and (d) $Sm_b$ ($L^*=30$, $M^*\simeq5.477$, $\beta P v_p=4.50$). In the left panels, the particles are colored according to the orientation of their long axis $\mathbf{\hat{u}}$ and in the middle panels according to the orientation of their short axis $\mathbf{\hat{w}}$ (cfr. Fig.~1 of the main text). Colors are defined according to the three axes of the simulation box (red, green, blue segments).  Diffraction patterns are calculated in the plane defined by $ \frac{2\pi} {\mathbf{n_\mathbf{u}}} \simeq \frac{2\pi} {z} $ and $ \frac{2\pi} {\mathbf{n_\mathbf{v}}} \simeq \frac{2\pi} {x} $ (i.e., the reciprocal of the ``blue''-``red'' axis shown in the snapshots). In the positionally-ordered smectic phase the sequence of bright dots is along the (reciprocal) main nematic director (corresponding to the reciprocal ``blue'' axis).  }
\label{SIfig1}
\end{figure*}

\begin{figure*}[h!t]
\begin{center}
\includegraphics[width=0.8\textwidth]{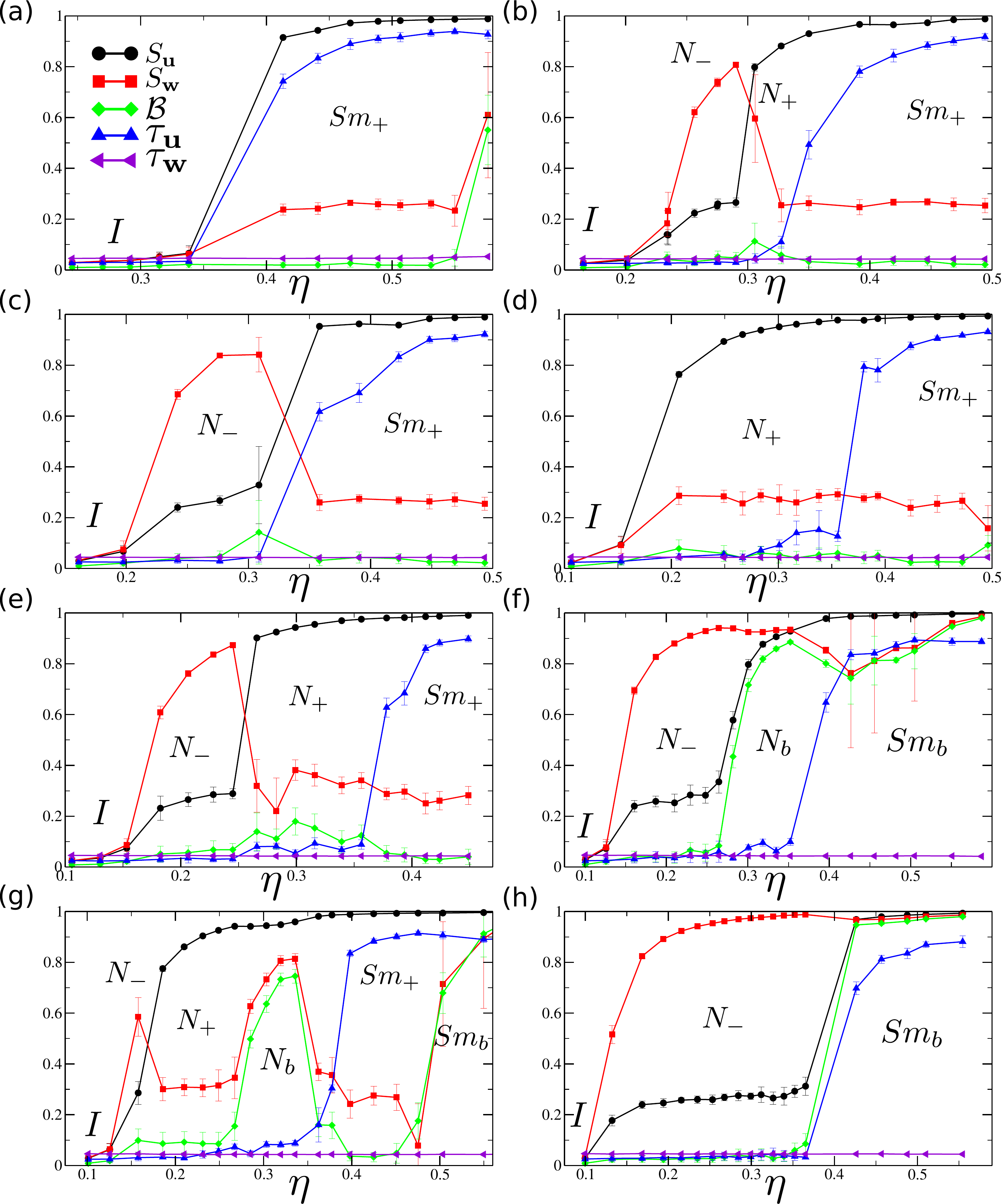}
\end{center}
\caption{Order parameters as a function of packing fraction $\eta$ for hard cuboids obtained by MC-$NPT$ simulations. Symbols correspond to simulation data and bars indicate standard deviation. The keys are the same for all the graphs. (a) $L^*=10$, $M^*\simeq3.16$ ($\nu=0$), (b) $L^*=16$, $M^*=4$ ($\nu=0$), (c) $L^*\simeq18.06$, $M^*=4.25$ ($\nu=0$), (d) $L^*=25$, $M^*=4.25$ ($\nu\simeq 0.065$), (e) $L^*=25$, $M^*=4.75$ ($\nu \simeq 0.02$), (f) $L^*=30$, $M^*\simeq5.47$ ($\nu=0$), (g) $L^*=30$, $M^*=5$ ($\nu\simeq0.03$), and (h) $L^*=30$, $M^*=7$ ($\nu\simeq-0.09$). Notice that we define $\nu=S/M-M/L$.}
\label{SIfig2}
\end{figure*}

\begin{figure*}[h!t]
\begin{center}
\includegraphics[width=0.9\textwidth]{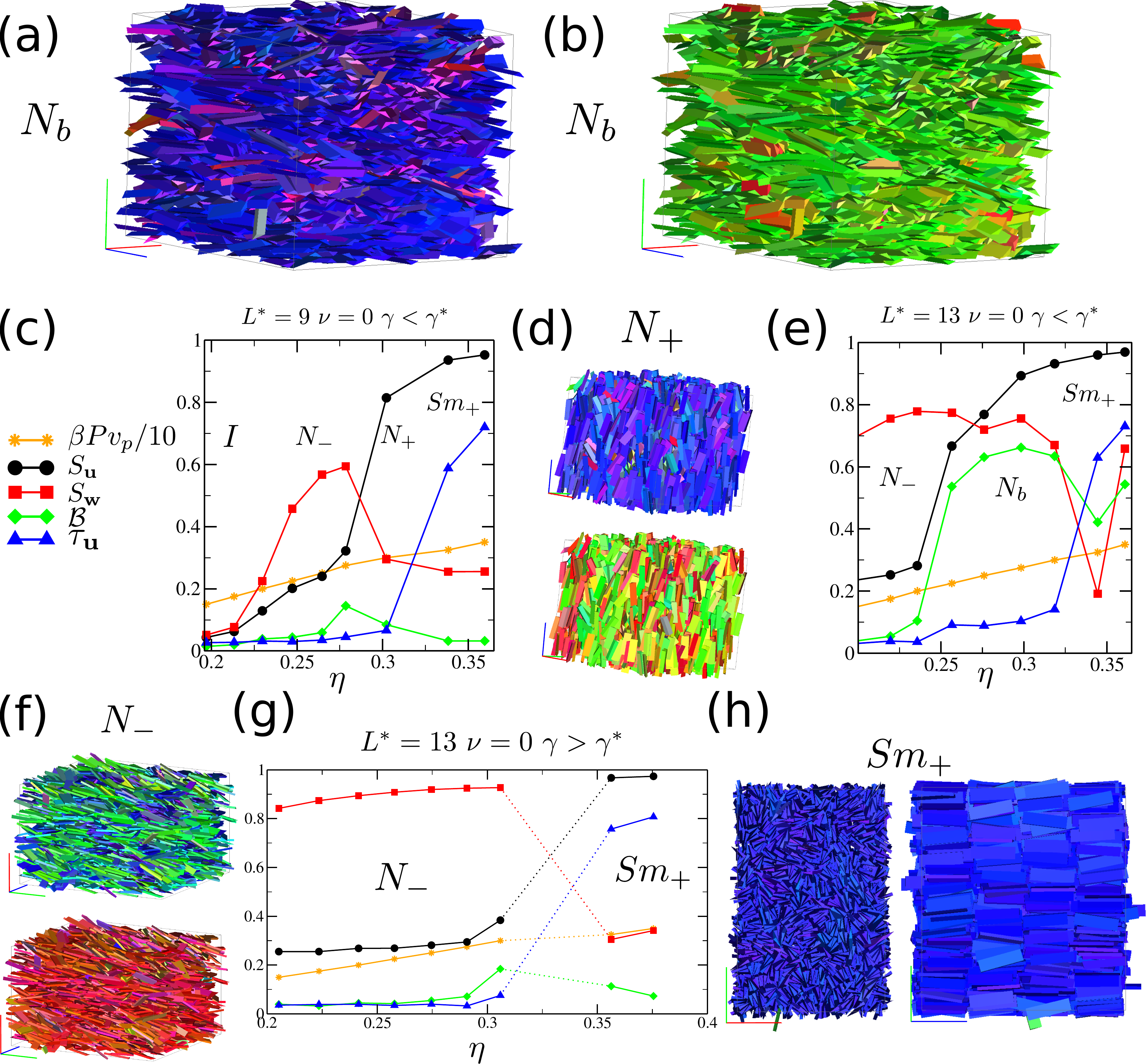}
\end{center}
\caption{(a-b) Representative snapshots of $N_b$ formed by triangular rods with $L^*=13$, $\nu=0$ and $\gamma<\gamma^*=\pi/3$. Same configuration is shown with particles colored according to the orientation of the long axis (a) or short axis (b). (c) Order parameters and equation of state ($\beta P v_p$, with $\beta=1/k_BT$ and $v_p$ the single-particle volume, as a function of packing fraction $\eta$), for triangular rods with $L^*=9$, $\nu=0$ and $\gamma<\gamma^*$. (d) Representative configuration of $N_+$ formed by triangular rods ($L^*=9$, $\nu=0$, $\gamma<\gamma^*$, $\beta P v_p=3.00$), color-coded according with orientation of the long (top) and short (bottom) axis. (e) Same as in (c) for $L^*=13$. (f) Same as in (d) for $N_-$ ($L^*=13$, $\nu=0$, $\gamma>\gamma^*$, $\beta P v_p=1.50$). (g) Same as in (e) for $\gamma>\gamma^*$. The first-order $N_- - Sm_+$ transition is indicated with dotted lines. (h) Representative configuration of $Sm_+$ formed by triangular rods with $L^*=13$, $\nu=0$, $\gamma>\gamma^*$, $\beta P v_p=3.50$; particles are colored according to the orientation of the long axis, two cross-sections are shown. }
\label{SIfig3}
\end{figure*}

\begin{figure*}[h!t]
\begin{center}
\includegraphics[width=0.9\textwidth]{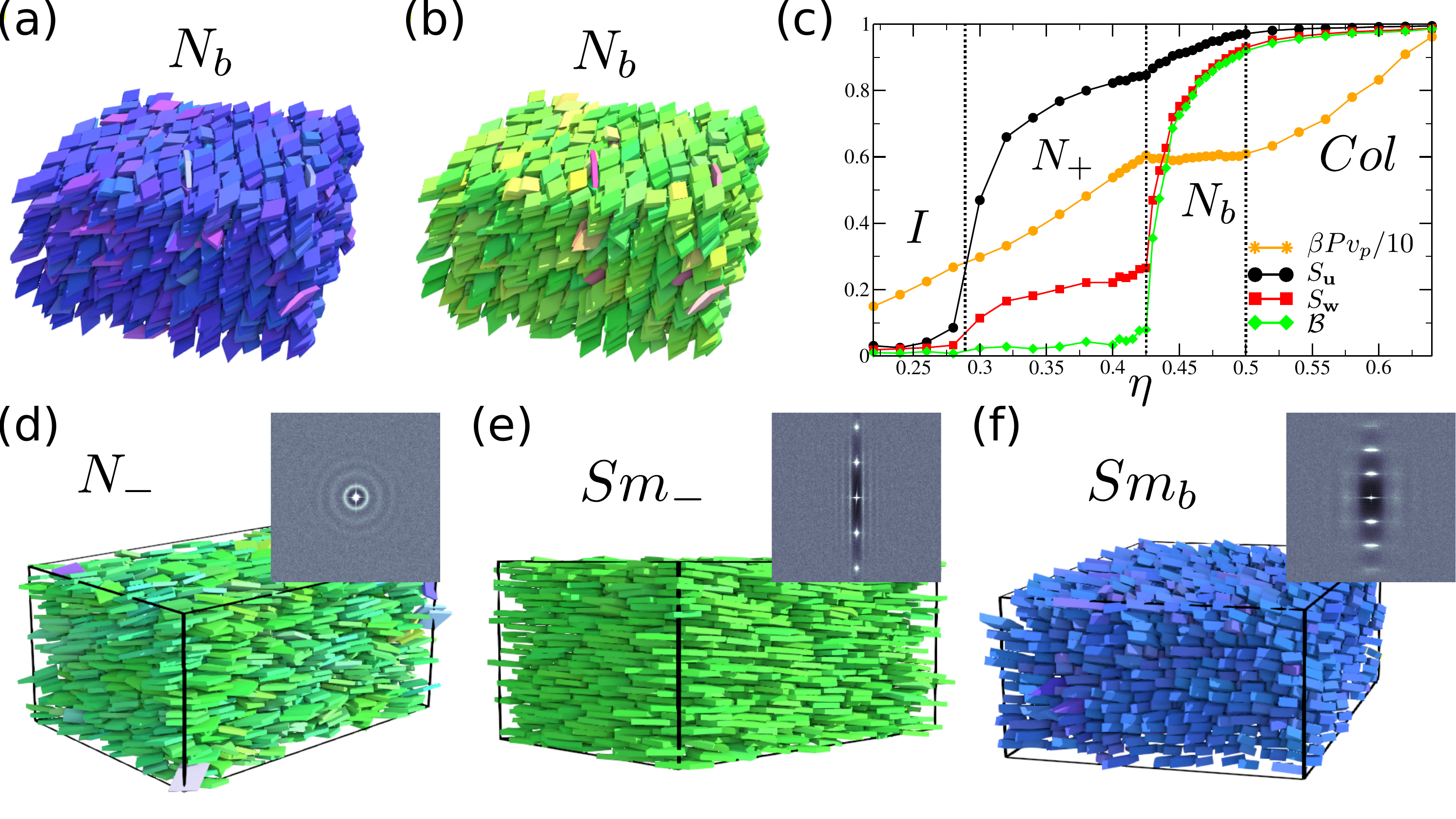}
\end{center}
\caption{(a-b) Representative snapshots of $N_b$ phase formed by rhombic particles with $L^*=11$ and $M^*=4$. The same configuration is shown with particles colored according to the orientation of their long axis (a) and their short axis (b).  (c) Order parameters as a function of packing fraction $\eta$ and equation of state for rhombic platelets with $L^*=11$ and $M^*=4$ as obtained from EDMD simulations. An isotropic to prolate nematic to biaxial nematic to columnar phase sequence is observed. Approximate boundaries are shown as dotted lines. (d) Representative configuration and corresponding diffraction pattern of $N_-$ phase of rhombic platelets with $L^*=11$ and $W^*=7$ ($\eta=0.42$). (e) $Sm_-$ phase for $L^*=11$ and $W^*=7$ ($\eta=0.52$). (f) $Sm_b$ for $L^*=11$ and $W^*=2$ ($\eta=0.52$).}
\label{SIfig4}
\end{figure*}

Representative configurations, diffraction patterns, and trends for the order parameters of the different models are shown in Figs.~\ref{SIfig1},~\ref{SIfig2},~\ref{SIfig3},~\ref{SIfig4}.

\section{Theory}

In this section, we describe our theoretical techniques and show some additional results for hard cuboids. In density functional theory, we express the free energy as a functional of the single-particle density $\rho(\mathbf{r},\Omega)$. We assume that the single-particle density has no spatial dependence, i.e. $\rho(\mathbf{r},\Omega) = \rho \psi(\Omega)$, where $\rho = N/V$ is the average density in a system of $N$ particles and volume $V$, and $\psi(\Omega)$ is the probability to find a particle with orientation $\psi(\Omega)$ in the interval $d\Omega$. The orientation of rigid, biaxial particles can be given by three Euler angles $\Omega = (\alpha, \beta, \gamma)$, with an integration measure $\int d\Omega = \int_0^{2\pi} d\alpha  \int_0^{\pi} \sin\beta d\beta \int_0^{2\pi} d\gamma = 8\pi^2$. The free energy density can be written as
	% ---------------
	\begin{eqnarray}\label{eq:freeEnergy}
	    \frac{\beta F\left[\psi(\Omega)\right]}{V} = 
	        \rho(\ln\mathcal{V}\rho-1)+\rho\int d\Omega\,\psi(\Omega)\ln\psi(\Omega)\nonumber\\
	         + \rho^2 B_2 + \frac{\rho^3}{2} B_3 + \ldots,
	\end{eqnarray}
	% ---------------
where $\beta = 1/(k_B T)$ is the inverse thermal energy and $\mathcal{V}$ is an irrelevant thermal volume factor. In the second-virial approximation, we truncate the excess free energy at $B_2$ and similarly in the third-virial theory we truncate at $B_3$. The second-virial term is
	% ---------------
	\begin{equation}\label{eq:B2}
	    B_2  = \frac{1}{2}\, \int d\Omega_1 \int d\Omega_2 E(\Omega_{12})\psi(\Omega_1)\psi(\Omega_2),
	\end{equation}
	% ---------------
where $\Omega_{12}=\Omega_2^{-1}\Omega_1$ is the relative orientation between two particles with orientations $\Omega_1$ and $\Omega_2$. The excluded volume $E(\Omega)$ in Eq.~\eqref{eq:B2} is defined as
	% ---------------
	\begin{eqnarray}\label{eq:exclVol2}
	    E(\Omega_{12}) &=& -\int d\mathbf{r}_{12} \, f(\mathbf{r}_{12},\Omega_{12}) \nonumber\\
	    &=& -\int d\mathbf{r}_{12}\, (\exp\left[-\beta U(\mathbf{r}_{12},\Omega_{12})\right] - 1),
	\end{eqnarray}
	% ---------------
where $f(\mathbf{r}_{12},\Omega_{12})$ is the Mayer function, $U(\mathbf{r}_{12},\Omega_{12})$ is the pair potential, and $\mathbf{r}_{12} = \mathbf{r}_2 -\mathbf{r}_1$ is the vector connecting the centers of the two particles. For hard particles we assume the pair potential to be
	% ---------------
	\begin{equation}
	\label{eq:potential}
	 \beta  U(\mathbf{r}_{12},\Omega_{12})= \left\{
	     \begin{array}{cl}
	       \infty, & \text{1 and 2 overlap;}  \\
	       0, & \text{otherwise} .
	     \end{array}
	   \right. 
	\end{equation}
	% ---------------
For hard cuboids, an analytic expression for $E(\Omega)$ is known~\cite{mulder2005}. The third-virial term is 
	% ---------------
	\begin{eqnarray}\label{eq:B3}
	    B_3 &=& \ddfrac{1}{3}\, \int d\Omega_1 \int d\Omega_2 \int d\Omega_3 \hat{E}(\Omega_{12},\Omega_{13})\psi(\Omega_1)\psi(\Omega_2)\psi(\Omega_3),\nonumber
	\end{eqnarray}
	% ---------------
with
	% ---------------
	\begin{eqnarray}\label{eq:exclVol3}
	    \hat{E}(\Omega_{12},\Omega_{13})  = -\int &d\mathbf{r}_{12}& \, \int d\mathbf{r}_{13} \left[  f(\mathbf{r}_{12},\Omega_{12}) f(\mathbf{r}_{13},\Omega_{13})\right. \nonumber\\
	    &\times& \left. f(\mathbf{r}_{13}-\mathbf{r}_{12},\Omega_{12}^{-1}\Omega_{13}) \right].
	\end{eqnarray}
	% ---------------

% \begin{table}
%   \begin{tabular*}{.5\linewidth}{@{\extracolsep{\fill}}| c || c | c | c | }
%     \hline
%     i & L & M & S \\
%     \hline
%     \hline
%     1 & x & y & z \\ \hline
%     2 & z & x & y \\ \hline
%     3 & y & z & x \\ \hline
%     4 & x & z & y \\ \hline
%     5 & y & x & z \\ \hline
%     6 & z & y & x \\ 
%     \hline
%   \end{tabular*}
%   \caption{Zwanzig orientations}\label{tab:zwanzig}
% \end{table}

First, we consider using the Zwanzig model, where we approximate the orientation distribution function of the six discrete, orthogonal orientations as $\psi(\Omega)=\psi_i$ with $i=1,\ldots,6$. Following Ref.~\cite{belli2011}, we can define the orientation vectors
	% ---------------
	\begin{eqnarray}\label{eq:zwanzigVect}
		\mathbf{X} &=& (L,M,S,L,M,S),\nonumber\\
		\mathbf{Y} &=& (M,S,L,S,L,M),\\
		\mathbf{Z} &=& (S,L,M,M,S,L),\nonumber
	\end{eqnarray}
	% ---------------
such that the dimensions of a particle with orientation $i$ in the $\hat{x},\hat{y},\hat{z}$ directions are $X_i,Y_i,Z_i$, respectively. Now the excluded volume [Eq.~\eqref{eq:exclVol2}] of two particles with orientations $i$ and $j$ is simply given by~\cite{belli2011}
	% ---------------
	\begin{equation}\label{eq:zwanzigE2}
		E_{ij}=(X_i+X_j)(Y_i+Y_j)(Z_i+Z_j).
	\end{equation}
	% ---------------
Similarly, we can write the three-particle excluded volume [Eq.~\eqref{eq:exclVol3}] for our Zwanzig model as
	% ---------------
	\begin{eqnarray}\label{eq:zwanzigE3}
		\hat{E}_{ijk} &=&  (X_i X_j + X_j X_k + X_i X_k) \\
		&\times& (Y_i Y_j + Y_j Y_k + Y_i Y_k) (Z_i Z_j + Z_j Z_k + Z_i Z_k).\nonumber
	\end{eqnarray}
	% ---------------
Using Eqs.~\eqref{eq:zwanzigVect}-\eqref{eq:zwanzigE3}, and minimizing the free energy [Eq.~\eqref{eq:freeEnergy} with appropriate replacements of $\int d\Omega \to \sum_{i=1}^6$] with respect to $\psi_i$ at fixed $\rho$ with the normalization condition $\sum_{i=1}^6 \psi_i = 1$ gives an Euler-Lagrange equation which can be solved iteratively for the equilibrium distribution $\psi_i^\text{eq}$. Then $\psi_i^\text{eq}$ can be used to identify the phase and to obtain the equilibrium free energy.

In our second model, the orientations are continuous rather than discrete and we instead consider expanding all Euler angle dependences in a complete basis of Wigner matrices $\D^l_{mn}(\Omega)$. For the excluded volume this gives
	% ---------------
	\begin{equation}\label{eq:wignerE}
	    E(\Omega_{12})=\sum_{l=0}^\infty\sum_{m,n=-l}^l  E^l_{mn} \D^l_{mn}(\Omega_{12}),
	\end{equation}
	% ---------------
where we can use the orthogonality of the Wigner matrices to write the coefficients as
	% ---------------
	\begin{equation}\label{eq:Elmn}
	    E^l_{mn} = \frac{2l+1}{8\pi^2} \int d\Omega \, E(\Omega)\, \D^l_{mn}(\Omega)^*.
	\end{equation}
	% ---------------
For $\psi(\Omega)$ we expand
	% ---------------
	\begin{equation}\label{eq:wignerPsi}
	        \psi( \Omega)= \sum_{l=0}^\infty \frac{2l+1}{8\pi^2} \sum_{m,n=-l}^l \langle \D^l_{mn} \rangle^* \D^l_{mn}( \Omega),
	\end{equation}
	% ---------------
where the coefficients $\langle \D^l_{mn} \rangle$ are order parameters, since they are given by
	% ---------------
	\begin{equation}\label{eq:orderPar}
	    \langle \D^l_{mn} \rangle = \int  d\Omega \, \D^l_{mn}(\Omega) \psi({\Omega}).
	\end{equation}
	% ---------------
	
 We can also choose to expand the logarithm of the orientation distribution function as
	% ---------------
	\begin{equation}\label{eq:wignerPsiExp}
	    \psi(\Omega)=\frac{1}{Z}\exp\left[\sum_{l=0}^\infty \, \sum_{m,n=-l}^l \psi^l_{mn} \D^l_{mn}(\Omega)\right],
	\end{equation}
	% ---------------
with the normalization of $\psi(\Omega)$ assured by the factor
	% ---------------
	\begin{equation}\label{eq:z}
	    Z = \int d\Omega \, \exp\left[\sum_{l=0}^\infty \, \sum_{m,n=-l}^l \psi^l_{mn} \D^l_{mn}(\Omega)\right].
	\end{equation}
	% ---------------
We prefer the expansion Eq.~\eqref{eq:wignerPsiExp}, since the coefficients $\psi^l_{mn}$ are unbounded and this expansion is expected to converge faster than Eq.~\eqref{eq:wignerPsi}. The Euler-Lagrange equation for the second-virial theory is then
	% ---------------
	\begin{equation}\label{eq:EL3}
		\psi^{l}_{mn}  = - \rho \,  \sum_{p=-l}^{l} E^{l}_{pn} \langle \D^{l}_{mp} \rangle^* 
	\end{equation}
	% ---------------
which together with Eq.~\eqref{eq:orderPar} can be solved for the set of coefficients $\psi^l_{mn}$, where the expansion in Eq.~\eqref{eq:wignerPsiExp} is truncated at some $l=l_\text{max}$. Based on the particle and phase symmetries, the number of $\psi^l_{mn}$ coefficients can be reduced to those with even $l,m,n$~\cite{mulder1989} and in addition, since $\psi(\Omega)$ is real we find that $\psi^l_{mn} = (-1)^{m-n}\psi^l_{-m-n}$. Here we focus on the coefficients with $l=2$, which are the only ones required by symmetry~\cite{mulder1989} and also the most important ones close to the dual shape where the isotropic-nematic transition is weakly first order. Of course, at higher densities we expect this approximation to be quantitatively inaccurate and the higher order (even) $l$ coefficients to be important.

For the full-orientation third-virial theory, we also expand 
	% ---------------
	\begin{equation}
	    \hat{E}(\Omega_{12},\Omega_{13}) = \sum_{l,m,n} \, \sum_{l',m',n'} \hat{E}^{ll'}_{mm'nn'} \D^l_{mn}(\Omega_{12}) \D^{l'}_{m'n'}(\Omega_{13}),
	\end{equation}
	% ---------------
where for brevity we write $\sum_{lmn}=\sum_{l=0}^\infty\sum_{m,n=-l}^l$, and where the coefficients are
	% ---------------
	\begin{eqnarray}
	        \hat{E}^{ll'}_{mm'nn'} = \frac{2l+1}{8\pi^2}&\,&\frac{2l'+1}{8\pi^2} \int \, d\Omega_{12} \int d\Omega_{13} \hat{E}(\Omega_{12},\Omega_{13})\nonumber\\
	         &\times& \D^l_{mn}(\Omega_{12})^*\D^{l'}_{m'n'}(\Omega_{13})^*.
	\end{eqnarray}
	% ---------------
 We calculate $\hat{E}^{ll'}_{mm'nn'}$ using Monte Carlo integration, with either 100 or 200 independent runs of with $10^{10}$ MC steps~\cite{belli2014,dussi2015}. The third-virial Euler-Lagrange equation is
	% ---------------
	\begin{eqnarray}
	        \psi^{l}_{mn}  &=& - \rho \,  \sum_{p=-l}^{l} E^{l}_{pn} \langle \D^{l}_{mp} \rangle^* 
	        -\frac{\rho^2}{2} \sum_{\tilde{l}\tilde{m}\tilde{n}\tilde{p}} \, \sum_{l'm'n'p'} \hat{E}^{\tilde{l}l'}_{\tilde{m}m'\tilde{n}n'}   \\ 
	        &\times& C(\tilde{l},\tilde{p};l',p';l,m) \, C(\tilde{l},\tilde{n};l',n';l,n) \,  \langle \D^{\tilde{l}}_{\tilde{p}\tilde{m}}\rangle^*  \langle \D^{l'}_{p'm'}\rangle^* ,\nonumber
	\end{eqnarray}
	% ---------------
where $C$ is the Clebsch-Gordan coefficient that arises from integrals over three Wigner matrices. Once the second or third-virial Euler-Lagrange equation is solved for the equilibrium $\{\psi^l_{mn}\}$, these can be used to obtain the order parameters [Eq.~\eqref{eq:orderPar}] and the free energy [Eq.~\eqref{eq:freeEnergy}].

Following the convention of Ref.~\cite{rosso2007}, we define four order parameters [which are proportional to those in Eq.~\eqref{eq:orderPar}], all of which are zero in the isotropic phase. In a uniaxial phase, the order parameters $S$ and $U$ are nonzero and $P=0=F$, with $S<0$ corresponding to a oblate nematic $N_-$, and $S>0$ corresponding to a prolate nematic $N_+$. In a biaxial nematic, all four of these order parameters are nonzero.

\begin{figure*}[h!t]
\begin{center}
\includegraphics[width=0.9\textwidth]{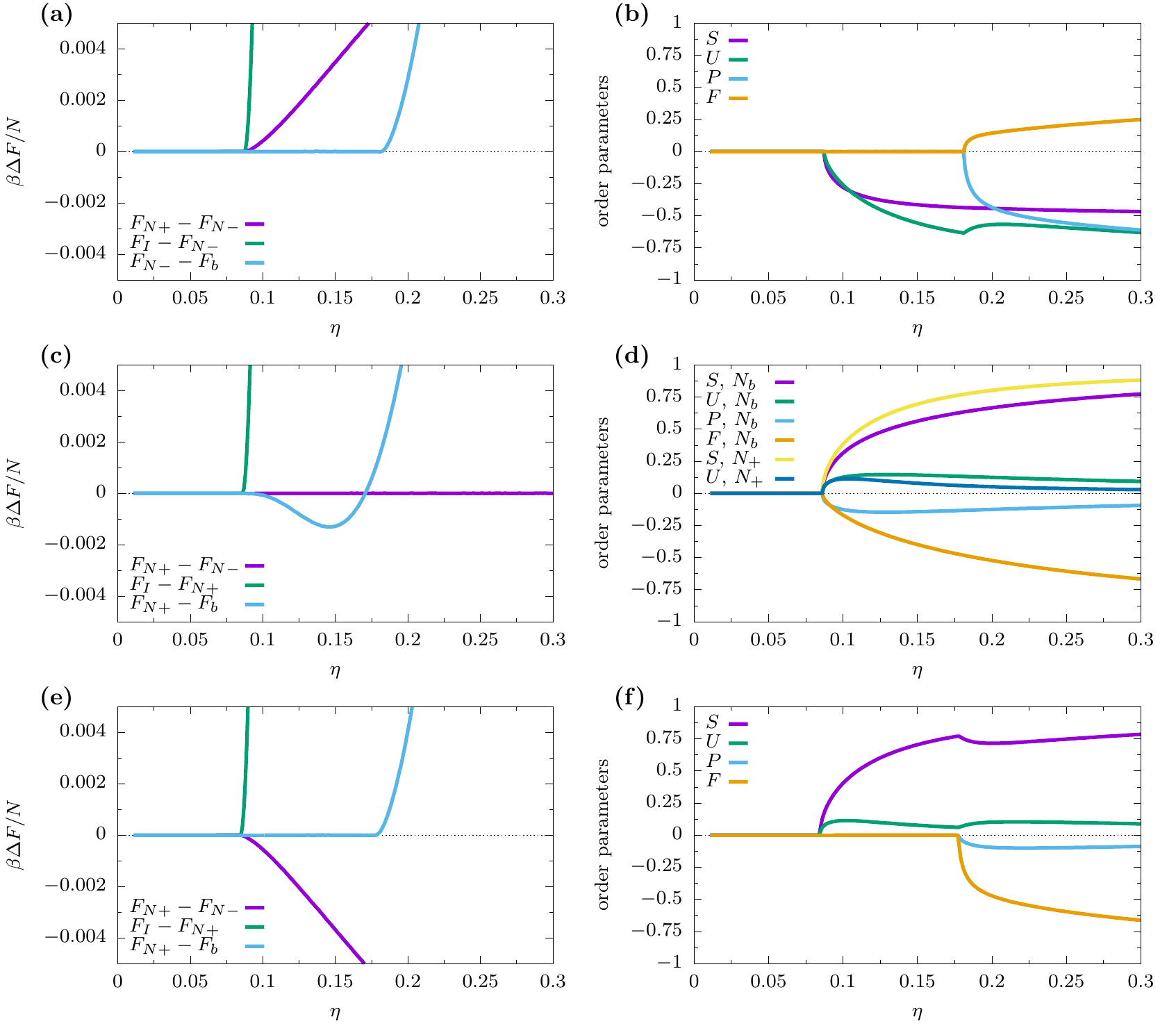}
\end{center}
\caption{Full-orientation second-virial theory results for cuboids with $M^*=8$ and $L^*=63$ (a-b), $L^*=64$ (c-d), and $L^*=65$ (e-f). The left column (a,c,e) shows the free energy difference $\Delta F$ between the different phases as a function of packing fraction $\eta$. The right column (b,d,f) shows the order parameters as a function of $\eta$ for the (possibly metastable) biaxial phase (order parameters $S,U,P,F$ are nonzero), oblate nematic ($P=0=F$, $S<0$), or prolate nematic ($P=0=F$, $S>0$).}
\label{SIfig:theory1}
\end{figure*}

In Fig.~\ref{SIfig:theory1}, we show the free energy differences [(a),(c),(e)] between the phases and the order parameters [(b),(d),(f)] as a function of packing fraction $\eta = \rho v_p$ using the full second-virial theory for three shapes with $M^*=8$ and: $L^*=63$ (a-b), $L^*=64$ (c-d), and $L^*=65$ (e-f). The plot of the free energy difference between the uniaxial and biaxial phases at $\nu=0$ ($L^*=64$) shows that the biaxial phase has a higher free energy than the uniaxial nematic (oblate or prolate, since these have identical free energies) for a small range of packing fractions above the isotropic phase, which corresponds to the dotted line at $\nu=0$ in Fig.~4(c) of the main text. Note that the order parameters shown in Fig.~\ref{SIfig:theory1}(d) are for both the biaxial nematic $N_b$, which is metastable in the region $0.1 \lesssim \eta \lesssim 0.17$, and for the prolate nematic $N_+$. We found that there was no direct isotropic-biaxial nematic transition for $M^*=8$, even if the Wigner matrix expansion was truncated at $l=4$ or $l=6$ (not shown). We also note that the isotropic-nematic coexistence region is extremely small for all shapes in the main text phase diagram Fig.~4(c), which is to be expected around the dual shape, though this is perhaps also underestimated by the truncation at $l=2$.

\begin{figure*}[h!t]
\begin{center}
\includegraphics[width=0.9\textwidth]{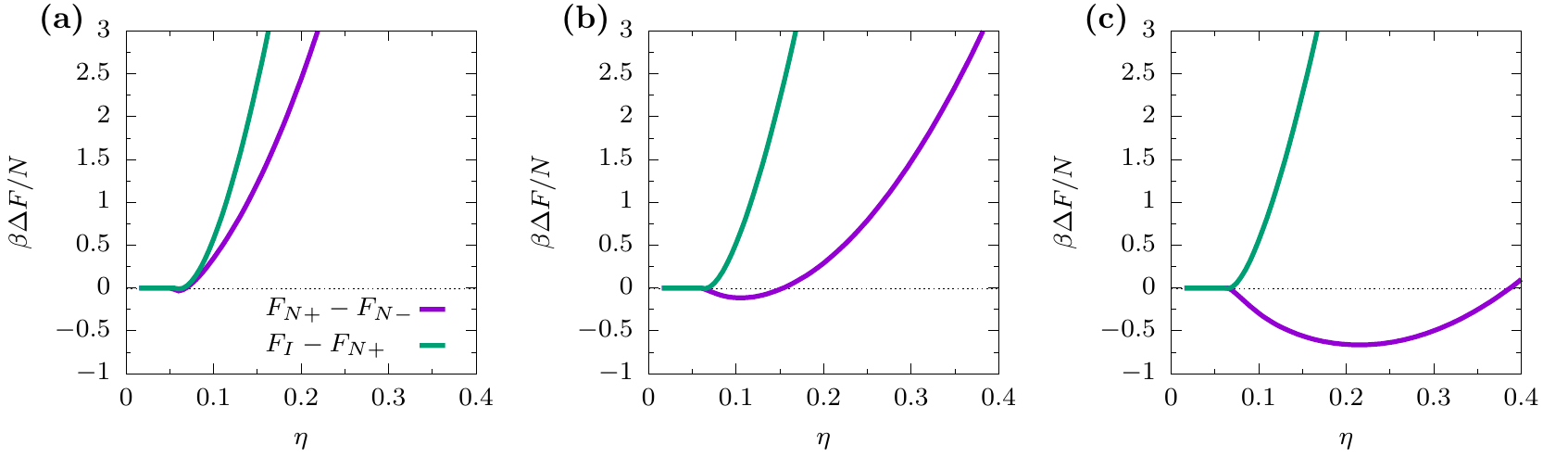}
\end{center}
\caption{Full-orientation third-virial theory results for cuboids with $L^*=64$. The free energy difference $\Delta F$ between the different phases as a function of packing fraction $\eta$ is shown for (a) $M^*=8$ ($\nu=0$), (b) $M^*=4.2$, and (c) $M^*=3.2$. Key applies to (a-c).}
\label{SIfig:theory2}
\end{figure*}

We make similar plots of the the free energy differences between phases as a function of packing fraction $\eta$ for our full-orientation third-virial theory in Fig.~\ref{SIfig:theory2}, for three shapes from main text Fig.~4(d). Here we see that for the dual shape [Fig.~\ref{SIfig:theory2}(a)], the oblate nematic is always favored over the prolate. For a more rod-like cuboid [Fig.~\ref{SIfig:theory2}(b)], the prolate is favored at low packing fraction and the oblate at higher packing fractions. For an even more rod-like cuboid [Fig.~\ref{SIfig:theory2}(c)], the prolate nematic is favored for a larger range of densities. The free energies in Fig.~\ref{SIfig:theory2} are typical of all shapes we studied, and in the main text Fig.~4(d)
 we choose (somewhat arbitrarily) to label the nematic phase sequences up to $\eta=0.4$ for long particles ($L \geq 50$) and up to $\eta=0.6$ for short particles ($L < 50$). However, we have not studied the stability of the nematic phases with respect to positionally ordered phases.
We also found that the biaxial nematic phase shifts to higher densities (for Fig.~\ref{SIfig:theory2}(a) $\eta \approx 0.3$, not shown) compared to what we found for the second-virial theory, and this phase has very small biaxial order parameters $P$, $F \sim 0.01$ (not shown). However, we caution that at these high densities our theory is not quantitatively accurate, both because we only take the $l=2$ term in the Wigner expansion and because the third virial term dominates over the second, and so we focus our attention on the uniaxial nematic behavior close to the isotropic-nematic transition. For all $L^*$ studied, we found that at $\nu=0$ the oblate nematic is preferred over the prolate within the full third-virial theory.

We also looked at the importance of the third-virial term as a function of particle aspect ratio. For spherocylinders in the isotropic phase, the ratio of the third-virial term to the second squared $B_{3,\text{iso}}/B_{2,\text{iso}}^2 \approx 0.3$ for short spherocylinders ($L/D = 10$) and less than $0.07$ for long spherocylinders ($L/D = 100$)~\cite{frenkel1987,frenkel1987erratum}. For dual-shaped cuboids (with $L^* = M^{*2}$), the same ratio between the virial terms is larger than $0.45$ for $L^*=10$ and about $0.25$ for $L^*=100$. Clearly, for the aspect-ratios studied here the third-virial term cannot be safely neglected, and perhaps even the higher virial terms should be considered.

\bibliography{biaxial_refs}

\pagebreak

\end{document}